\documentclass{iopconfser}
\usepackage{cite}
\usepackage{graphicx}
\usepackage[rightcaption]{sidecap}

\newcommand{\dd}{{\rm{d}}}
\newcommand{\PP}{{\mathcal{P}}}   
\newcommand{\QQ}{{\mathcal{Q}}}   
\newcommand{\scri}{{\mathcal{I}}} 
\newcommand{\rovno}{\!\!\!\!& = &\!\!\!\!}

\def\scrisub{{_{\!\mathcal{I}}}}

\begin{document}

\title{Various metric forms of all type D black holes\\ and their application}

\author{Ji\v{r}\'{i} Podolsk\'{y}}
\vspace{0mm}

\affil{Institute of Theoretical Physics, Charles University, Prague, Czechia}
\vspace{1mm}

\email{jiri.podolsky@matfyz.cuni.cz}
\vspace{3mm}

\begin{abstract}
Several recent results concerning the complete class of exact solutions to the Einstein--Maxwell--$\Lambda$ equations of type D with a double-aligned electromagnetic field are summarized. This large class of spacetimes includes charged black holes with rotation, NUT parameter, and acceleration. In particular, we present their Pleba\'{n}ski--Demia\'{n}ski, Griffiths--Podolsk\'{y}--Vr\'{a}tn\'{y}, and Astorino metric representations, we discuss mutual relations between them, and demonstrate their usefulness for investigation of physical and geometrical properties (singularities, horizons, conformal diagrams, ergoregions, rotating strings causing the acceleration, thermodynamics). It includes the proof that such black holes emit gravitational radiation if, and only if, they accelerate. Very recent extension to type~D black holes with non-aligned Maxwell field is also mentioned.
\end{abstract}

\section{Introduction}

This contribution is based on several joint papers with Adam Vr\'{a}tn\'{y} \cite{PV:2021, PV:2023, PV:2020}, Hryhorii Ovcharenko and Marco Astorino \cite{OPA:2025a, OPA:2025b}, Francisco Fern\'{a}ndez-\'{A}lvarez and Jos\'{e} Senovilla \cite{FPS:2024}, and Hryhorii Ovcharenko \cite{PodolskyOvcharenko:2025a, OvcharenkoPodolsky:2025b} --- which we published during the last few years. The common theme is the investigation of a family of exact spacetimes representing black holes of algebraic type D that solve the Einstein--Maxwell field equations, and also include any value of the cosmological constant $\Lambda$.

The first such solutions were found more than 100 years ago, namely the celebrated Schwarzschild (1916), charged Reissner--Nordstr\"{o}m (1916, 1918), and cosmological Kottler--Weyl--Trefftz (1918--1922) spacetimes. In the 1960s these were generalized by Kerr and Newman (1963, 1965) to include rotation and also the NUT (Newman--Unti--Tamburino) twist (1963). Possible uniform acceleration was added, after Kinnersley and Walker (1970) properly interpreted the $C$-metric (1918, 1962).

They all belong to a general family of (spherically/axially symmetric) spacetimes of the Weyl type~D, with (double-aligned, non-null) electromagnetic field and any~${\Lambda}$. It was discovered by Debever in 1971, but its more convenient form was presented in 1976 by Pleba\'nski and Demia\'nski \cite{PlebanskiDemianski:1976} (see \cite{Stephanietal:2003, GriffithsPodolsky:2009} for reviews). After 2005, this large Pleba\'nski--Demia\'nski (PD) class of spacetimes was investigated in more detail by Griffiths and Podolsk\'y (GP) \cite{GriffithsPodolsky:2005, GriffithsPodolsky:2006, PodolskyGriffiths:2006}, and a few years ago again by Podolsk\'y and Vr\'atn\'y (PV) \cite{PV:2021, PV:2023}.

Recently, these type~D spacetimes were reconsidered by Astorino~(A). He employed solution-generating techniques and obtained a novel metric form \cite{Astorino:2024b, Astorino:2024a}. Its very nice feature is that it directly allows limits to \emph{all} the subcases, including the peculiar and elusive accelerating black hole with the NUT parameter (and no Kerr rotation). This was previously thought to exist only outside the type~D class because we showed in \cite{PV:2020} that the Chng--Mann--Stelea solution \cite{ChngMannStelea:2006} is of algebraic type~I (see also \cite{Astorino:2023a, Astorino:2023b}).  After the presentation of the new A metric it was not obvious if it is fully equivalent to the PD metric.

This was the motivation of our work \cite{OPA:2025a}. In it, we put the A metric into a more compact A$^+$ form, and then elucidated its relations to PD, GP, and PV metric representations of this family of black hole solutions to the Einstein-Maxwell-$\Lambda$ theory. Due to considerable complexity of the exact relation between the various forms, in \cite{OPA:2025a} we restricted the analysis only to the case when  ${\Lambda=0}$. Extension to the most general situation with any value of the cosmological constant was performed in \cite{OPA:2025b}.
\newpage

\section{Pleba\'nski--Demia\'nski (PD) metric}

The metric of the complete class of PD spacetimes presented in \cite{PlebanskiDemianski:1976} is
\begin{equation}
\dd s^2=\frac{1}{(1-p'r')^2} \bigg[\!
 -\frac{Q'\,(\dd\tau'-p'^{\,2}\dd\sigma')^2}{r'^{\,2}+p'^{\,2}}
 +\frac{r'^{\,2}+p'^{\,2}}{Q'}\,\dd r'^{\,2}
 +\frac{r'^{\,2}+p'^{\,2}}{{P'}}\,\dd p'^{\,2}
 +\frac{P'\,(\dd\tau'+r'^{\,2}\dd\sigma')^2}{r'^{\,2}+p'^{\,2}} \bigg],
  \label{oldPDMetric}
\end{equation}
\vspace{-2mm}

\qquad \qquad \qquad
$P'= \,\,k' +2 n'p' - \epsilon'p'^{\,2} +2 m'p'^{\,3\,}-\,(k'+e'^2+g'^2+\Lambda'/3)\,p'^{\,4} $\,,
\vspace{2mm}

\qquad \qquad \qquad
$Q'=(k'+e'^2+g'^2) -2m'r' +\epsilon'r'^{\,2} -2n'r'^{\,3}-(k'+\Lambda'/3)\,r'^{\,4} $\,,
\vspace{3mm}

\noindent
which are simple quartic functions of $p'$ and $r'$, respectively. However, it is \emph{not clear what is the physical meaning} of the 7 free parameters $k'$, $n'$, $\epsilon'$, $m'$, $e'$, $g'$, $\Lambda'$ (the value of cosmological constant $\Lambda'$ is given).

\section{Griffiths--Podolsk\'y (GP) metric}

This problem was resolved in the works  \cite{GriffithsPodolsky:2005, GriffithsPodolsky:2006, PodolskyGriffiths:2006} (and reviewed in \cite{GriffithsPodolsky:2009}) by performing a suitable transformation of coordinates and by introducing new parameters with a direct physical meaning. The resulting GP form of the family of type D black holes reads
\begin{equation}
\dd s^2 = \frac{1}{\Omega^2}
  \bigg[\!-\frac{\QQ}{\rho^2}\big[\dd t- \big(\bar{a}\sin^2\theta +4\bar{l}\sin^2\!\frac{\theta}{2}\big)\dd\varphi \big]^2 \!
   + \frac{\rho^2}{\QQ}\,\dd \bar{r}^2
   + \frac{\rho^2}{\PP}\,\dd\theta^2
   + \frac{\PP}{\rho^2}\sin^2\theta \big[ \bar{a}\,\dd t -\big(\bar{r}^2+(\bar{a}+\bar{l})^2\big)\dd\varphi \big]^2
 \bigg], \label{GP-metric}
\end{equation}
where
${\,\Omega  = 1 - {\displaystyle\frac{\bar{\alpha}}{\omega}}\,
  ( \bar{l} + \bar{a} \cos\theta )\,\bar{r}\,}$,\quad
${\,\rho^2    = \bar{r}^2+( \bar{l}+ \bar{a} \cos \theta)^2\,}$, and
\vspace{-4mm}
\begin{equation}
\PP = 1-a_3\cos\theta-a_4\cos^2\theta\,,  \quad
\QQ = (\omega^2\bar{k}+\bar{e}^2+\bar{g}^2)-2\bar{m}\,\bar{r}+\bar{\epsilon}\,\bar{r}^2
       -2\bar{\alpha}\,\frac{\bar{n}}{\omega}\,\bar{r}^3
       -\Big(\bar{\alpha}^2\bar{k}+\frac{\Lambda}{3}\Big)\,\bar{r}^4\,.  \label{Q}
\end{equation}
This metric contains parameters with usual meaning, namely the Kerr rotation~$\bar{a}$ and the NUT parameter~$\bar{l}$ (both related to the twist $\omega$), acceleration $\bar{\alpha}$, mass $\bar{m}$,  electric and magnetic charges $\bar{e}$ and $\bar{g}$, and the cosmological constant~$\Lambda$. But there still remain five auxiliary constants $a_3, a_4, \bar{k}, \bar{\epsilon}, \bar{n}$ which have to be expressed in terms of these physical parameters. Such an explicit and unique relation
was found in \cite{GriffithsPodolsky:2005, GriffithsPodolsky:2006, PodolskyGriffiths:2006}, but it is \emph{quite complicated} (see equations (16.15)--(16.20)  in  \cite{GriffithsPodolsky:2009}). Nevertheless, this enabled us to obtain all previously known black holes in their standard (Boyer--Lindquist) forms as special subcases, and discuss the structure of the whole family. In fact, all famous black holes are obtained from the GP metric (\ref{GP-metric}), (\ref{Q}) by setting the corresponding new physical parameters to zero. For example, the Kerr-Newman-(anti-)de~Sitter black hole is recovered by simply setting ${\bar{l}=0, \bar{\alpha}=0, \omega=\bar{a}}$.

\section{Podolsk\'y--Vr\'atn\'y (PV) metric}

Further considerable improvement was achieved in \cite{PV:2021, PV:2023} by transforming\footnote{It involved specific modification of the mass and charge parameters, conformal rescaling, and useful gauge choice of~$\omega$.}  the GP form to
\begin{equation}
\dd s^2 = \frac{1}{\Omega^2}
  \bigg[\!-\frac{Q}{\rho^2}\big[\dd t- \big(\hat{a}\sin^2\theta +4\hat{l}\sin^2\!\frac{\theta}{2}\big)\dd\varphi \big]^2 \!
   + \frac{\rho^2}{Q}\,\dd \hat{r}^2
   + \frac{\rho^2}{P}\,\dd\theta^2
   + \frac{P}{\rho^2}\sin^2\theta \big[ \hat{a}\,\dd t -\big(\hat{r}^2+(\hat{a}+\hat{l}\,)^2\big)\dd\varphi \big]^2
 \bigg], \label{PV-metric}
\end{equation}
where
${\,\Omega  = 1 - {\displaystyle\frac{\hat{\alpha}\,\hat{a}}{\hat{a}^2+\hat{l}^{\,2}}}\,
  ( \hat{l} + \hat{a} \cos\theta )\,\hat{r}\,}$,\quad
${\,\rho^2    = \hat{r}^2+( \hat{l}+ \hat{a} \cos \theta)^2\,}$, and
\vspace{-2mm}
\begin{eqnarray}
P \rovno 1
   -2\,\Big(\,\frac{\hat{\alpha}\,\hat{a}}{\hat{a}^2+\hat{l}^{\,2}}\,\,\hat{m} - \frac{\Lambda}{3}\,\hat{l}\, \Big)(\hat{l}+\hat{a}\,\cos\theta) +\Big[\frac{\hat{\alpha}^2 \hat{a}^2}{(\hat{a}^2+\hat{l}^{\,2})^2} (\hat{a}^2-\hat{l}^{\,2} + \hat{e}^2 + \hat{g}^2)
       + \frac{\Lambda}{3} \Big](\hat{l}+\hat{a}\,\cos\theta)^2 \,,  \nonumber\\
Q  \rovno \Big[\,\hat{r}^2 - 2\hat{m}\, \hat{r}  + (\hat{a}^2-\hat{l}^{\,2}+\hat{e}^2+\hat{g}^2) \Big]\!
            \Big(1+\hat{\alpha}\,\hat{a}\,\frac{\hat{a}-\hat{l}}{\hat{a}^2+\hat{l}^{\,2}}\, \hat{r}\Big)\!
            \Big(1-\hat{\alpha}\,\hat{a}\,\frac{\hat{a}+\hat{l}}{\hat{a}^2+\hat{l}^{\,2}}\, \hat{r}\Big)\\
   &&  \hspace*{49mm}  - \frac{\Lambda}{3}\Big[\,
      (\hat{a}^2+3\,\hat{l}^{\,2})\,\hat{r}^2 + 2\hat{\alpha}\,\hat{a}\,\hat{l}\,\,\frac{\hat{a}^2-\hat{l}^{\,2}}{\hat{a}^2+
      \hat{l}^{\,2}}\,\hat{r}^3 + \hat{r}^4 \Big]. \nonumber
\end{eqnarray}
The main advantage of this PV form is that the metric is simple, with all the functions \emph{expressed directly} in terms of the physical parameters, namely the mass $\hat{m}$, electric and magnetic charges $\hat{e}$ and~$\hat{g}$, the Kerr rotation~$\hat{a}$, the NUT parameter $\hat{l}$, and acceleration $\hat{\alpha}$. Moreover, if the cosmological constant~$\Lambda$ vanishes, \emph{the metric functions $P$ and $Q$ are factorized}. This is very helpful for investigation of various physical and geometrical properties of this large family of rotating, charged, and accelerating black holes, namely their singularities, horizons, ergoregions, conformal infinities, cosmic strings, or thermodynamics \cite{PV:2021, PV:2023}.
\newpage

\noindent
Indeed, all four metric functions are directly related to the spacetime geometry:

$\bullet$ \ \ \emph{curvature singularity} is located at  ${\rho=0\,\,\Leftrightarrow\,\, \hat{r}=0 \hbox{\,\ and\,\ } \hat{l}+ \hat{a} \cos \theta = 0 }$\,,

$\bullet$ \ \ \emph{conformal infinity}~$\scri$ is identified by  ${\Omega=0}$\,,

$\bullet$ \ \ \emph{axes} at ${\theta=0}$ and ${\theta=\pi}$ are \emph{regular} if and only if  ${P=1}$\,,

$\bullet$ \ \ \emph{horizons} are located at  ${Q=0}$\,.

\begin{SCfigure}[1.1][ht!]
\includegraphics[width=0.45\textwidth]{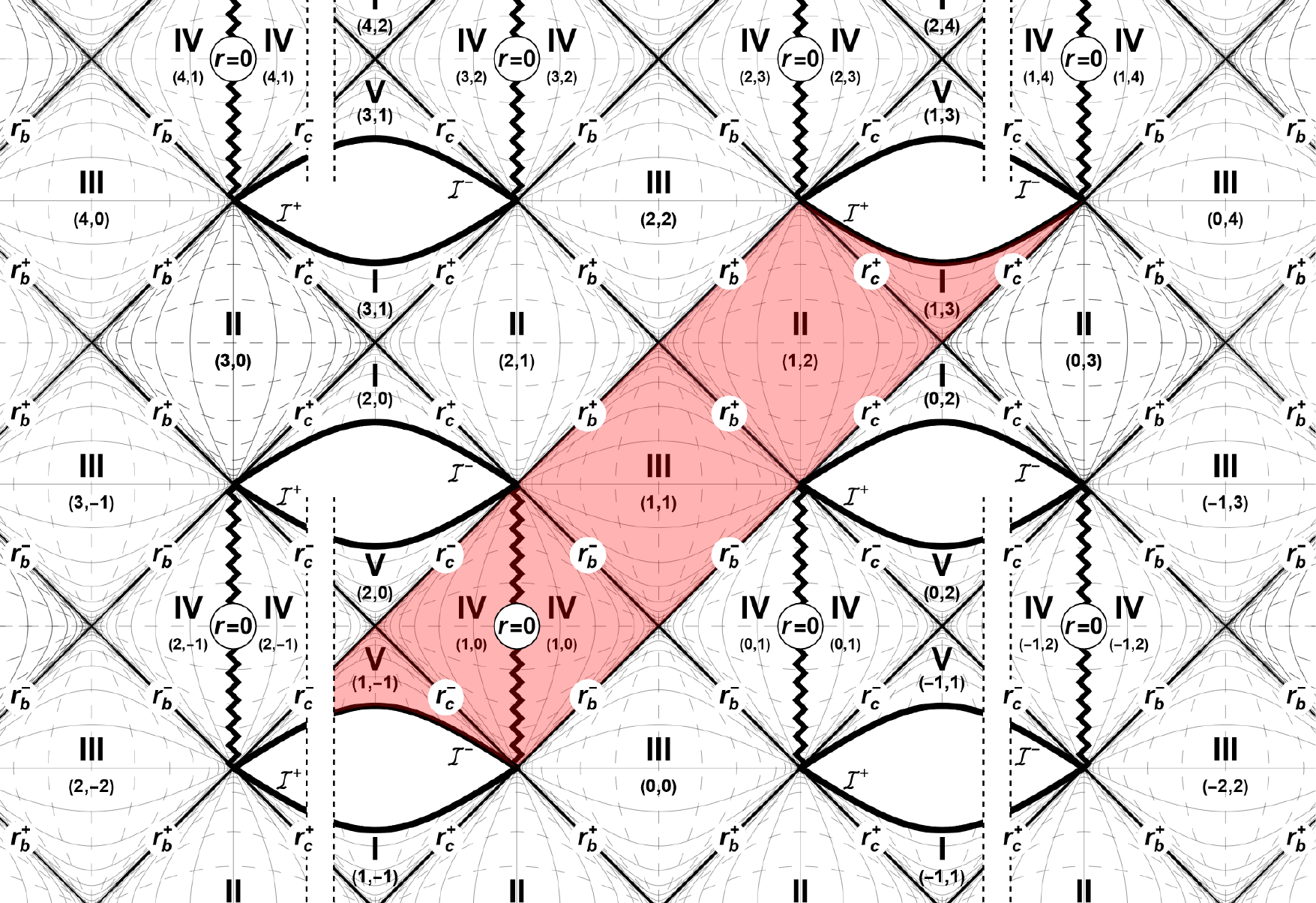}
\hspace{3mm}
\vspace{-8mm}
\caption{\small
Penrose conformal diagram of the completely extended PV spacetime showing the global structure of the family of accelerating and rotating charged NUT black holes of type D. Here we show a 2-dimensional section ${\theta, \varphi =\hbox{const.}}$ with the ring-like curvature singularity at ${\hat{r} = 0}$, ${\cos \theta = -\hat{l}/\hat{a}}$ (vertical zigzag lines). The vertical dashed parallel lines indicate a separation of (generally) distinct asymptotical regions close to~$\scri^\pm$.}
\label{Fig}
\end{SCfigure}

\noindent
In particular, generally there are \emph{four distinct horizons}, namely the inner/outer {\em black-hole horizons} at~$r_b^\pm$ and inner/outer {\em cosmo-acceleration horizons} at~$r_c^\pm$, naturally ordered as
${r_c^- < 0 < r_b^- < r_b^+ < r_c^+}$. In such a case, the Penrose conformal diagram representing the global structure has the form shown in figure~\ref{Fig}. It can be seen that in the whole universe there are  \emph{infinitely many black holes}.

\vspace{-1mm}
\vspace{4mm}

\section{Astorino (A) and Ovcharenko--Podolsk\'y--Astorino (A$^+$) metric}

In 2024, all type~D black holes were reconsidered again by Astorino using solution-generating techniques, and a novel~(A) metric form was thus obtained \cite{Astorino:2024b, Astorino:2024a}. Its advantage is that it directly admits limits to \emph{all the subcases}, including the peculiar accelerating solutions with (just) the NUT parameter, which  was previously considered to exist outside the type~D class \cite{ChngMannStelea:2006,PV:2020}, see also \cite{Astorino:2023a, Astorino:2023b}.

After the presentation of this new A metric it was not obvious whether it is equivalent to the PD metric. This motivated our joint works \cite{OPA:2025a, OPA:2025b} in which we put the A metric into an even \emph{more compact}, improved A$^+$ form. Then, we elucidated its explicit relations to the PD, GP, and PV metrics. In \cite{OPA:2025a} we restricted the analysis only to the case ${\Lambda=0}$, and subsequently we extended it to the most general situation with any $\Lambda$ \cite{OPA:2025b}.

Specifically, the A$^+$ metric simply reads
\begin{equation}
    \dd s^2=\frac{1}{\Omega^2}\bigg[\!
    - \frac{\Delta_r}{\rho^2}(A\,\dd t - B\,\dd\varphi)^2
    + \frac{\Delta_x}{\rho^2}(C\,\dd t + D\,\dd\varphi)^2
    + C_f\,\rho^2 \Big(\,\frac{\dd r^2}{\Delta_r}
    + \frac{\dd x^2}{\Delta_x}\,\Big)\bigg],
    \label{A+metric}
\end{equation}
where  ${\,\Omega  = 1-\alpha\, r\, x\,}$,\quad ${\,\rho^2  = AD+BC\,}$, \quad ${C_f=\hbox{const.}}$,
\begin{equation}
    A = 1 + \alpha^2(l^2-a^2)\,x^2,\ \quad
    B = a + 2l\,x + a\,x^2,\ \quad
    C = a + 2\alpha l\,r + \alpha^2 a\, r^2,\ \quad
    D = (l^2-a^2) + r^2,
\end{equation}
\vspace{-7mm}
\begin{eqnarray}
\Delta_r \rovno (1-\alpha^2r^2)\big[(r-m)^2 - (m^2+l^2-a^2-e^2-g^2)\big]
   - \frac{\Lambda}{3}\Big(\,\frac{3l^2}{1+\alpha^2 a^2}\,r^2
     + \frac{4\alpha a l}{1+\alpha^2 a^2}\,r^3 + r^4 \Big),\nonumber\\[0mm]
\Delta_x \rovno (1-x^2)\big[(1-\alpha m \,x)^2-\alpha^2x^2(m^2+l^2-a^2-e^2-g^2)\big] \\
   && \hspace{44mm} - \frac{\Lambda}{3}\Big(\,\frac{3l^2}{1+\alpha^2 a^2}\,x^2
    +\frac{4al}{1+\alpha^2 a^2}\,x^3 + \frac{a^2+\alpha^2(a^2-l^2)^2}{1+\alpha^2 a^2}\,x^4 \Big).\nonumber
\end{eqnarray}
The physical parameters, namely the mass $m$, electric and magnetic charges $e$ and~$g$, the Kerr rotation~$a$, the NUT parameter $l$, and acceleration $\alpha$ \emph{can be set to zero}. Interestingly, for ${a=0}$ the elusive class of type D accelerating NUT black holes is obtained (which could not be identified in the PV metric form (\ref{PV-metric}) because ${\hat{a}=0}$ removes the acceleration parameter $\hat{\alpha}$, cf. \cite{WuWu:2024}). For ${\Lambda=0}$ \emph{the metric functions $\Delta_r $ and~$\Delta_x $ are very simple, and factorized}. Their zeros readily identify the horizons and the axes, respectively.

Explicit relations between the new A/A$^+$ and old PD, GP, and PV coordinates and parameters are very complicated, see \cite{OPA:2025a, OPA:2025b}. Let us present here only the expression for the GP/PV \emph{acceleration parameter}
\begin{equation}
\bar{\alpha} \equiv \hat{\alpha} = \alpha\,\big[\,a - \alpha^2 a (a^2-l^2)
+ \sqrt{a^2+\alpha^4 a^2(a^2-l^2)^2+2\alpha^2(a^2-l^2)(a^2-2l^2)}\, \big] /(2 \omega)\,.
\end{equation}
Obviously, they \emph{are not the same}, but ${\,\alpha=0\,}$ implies ${\,\bar{\alpha}=0=\hat{\alpha}\,}$. Moreover, for {\em vanishing NUT} (${l=0}$), with the most natural choice of twist ${\omega=a\,}$,  we get ${\,\bar{\alpha}=\hat{\alpha}=\alpha}$. Also in many other subcases the new and old parameters agree, except when ${\Lambda a^2\ne0}$ or ${\Lambda l^2\ne0}$, and also when ${a=0}$ that gives ${\bar{a}\ne0}$, ${\bar{l}\ne0}$ \cite{OPA:2025a, OPA:2025b}.

\section{Gravitational radiation generated by accelerating black holes}

As an interesting application of these new forms of all type D black holes, we applied them to study the gravitational radiation they generate in spacetimes with  ${\Lambda>0}$ which have \emph{de~Sitter-like conformal infinity}. In \cite{FPS:2024} we employed the PV and A$^+$ forms to the  \emph{Fern\'andez-\'Alvarez--Senovilla criterion} developed recently \cite{FranJose:2020, FranJose:2022}. The general framework is based on evaluation of quantities in conformal spacetime, in particular the commutator of the canonical electric and magnetic parts of the re-scaled Weyl tensor at~$\scri$ which gives \emph{asymptotic super-Poynting vector}~$\mathbf{\overline{P}}$ (for more details see Jos\'{e} Senovilla's contribution to these proceedings \emph{Characterization of gravitational radiation at infinity with a cosmological constant}). After long calculation, we arrived at a surprisingly simple~formula
\begin{equation}
 \mathbf{\overline{P}} = \alpha\, \Big(\frac{3}{|\Lambda|}\Big)^{\!\frac{5}{2}}
     18(1+\alpha^2a^2)^3\,
     P_{\!\scrisub} Q_{\scrisub}\big(Q_{\scrisub}-\alpha^2 P_{\!\scrisub} \big) \,
     \frac{|\psi_2|^2}{\rho^6_\scrisub }
     \,\partial_{x}\,,
    \label{super-Poynting-scri}
\end{equation}
where $P_{\!\scrisub}(x), Q_{\scrisub}(x), \rho_\scrisub(x)$ are obtained from the functions $P, Q, \rho$ (or $\Delta_x, \Delta_r, \rho$), evaluated on $\scri$ given by ${\Omega=0}$, and~$\psi_2$ is the only non-zero scalar of the re-scaled Weyl tensor. It is now explicitly seen that ${\mathbf{\overline{P}} = 0 \Leftrightarrow  \alpha =0 }$. It means, that these \emph{black holes emit gravitational radiation~if} (\emph{and only if}) \emph{they accelerate}. This is a neat conclusion, which provides a strong test of the criterion, but also justifies the physical interpretation of the parameter $\alpha$ as acceleration. Moreover, it extends and clarifies previous investigation \cite{KrtousPodolsky:2004} of asymptotic directional structure of radiation: the super-Poynting vector~$\mathbf{\overline{P}}$ points into the \emph{maximum} of the local radiation pattern on the de~Sitter-like~$\scri$.

\section{Further extension: Black holes of type D with non-aligned electromagnetic field}

Finally, let us briefly mention our most recent result, which is the derivation of a large family of type~D black holes for which the Maxwell field is \emph{fully non-aligned} with the gravitational field. This is a completely new family of black holes that \emph{does not belong to the Pleba\'{n}ski--Demia\'{n}ski class}. Its first presentation was given by Hryhorii Ovcharenko at the A1 session of GR24/Amaldi16 on July 14, 2025. Our contribution \emph{A novel class of rotating black holes with non-aligned electromagnetic field} is given elsewhere in these proceedings. An extensive description can be found in \cite{OvcharenkoPodolsky:2025b}, together with the derivation of a very interesting novel subclass which represents Kerr black hole in a uniform magnetic field \cite{PodolskyOvcharenko:2025a}.

\section*{Acknowledgments}

This work was supported by the Czech Science Foundation Grant No.~GA\v{C}R 23-05914S and also by the Charles University Grant No.~GAUK 260325.




\begin{thebibliography}{10}


\bibitem{PV:2021}
Podolsk\'{y} J and Vr\'{a}tn\'{y} A
2021
New improved form of black holes of type D
\emph{Phys. Rev.} D {\bf 104} 084078

\bibitem{PV:2023}
Podolsk\'{y} J and Vr\'{a}tn\'{y} A
2023
New form of all black holes of type D with a cosmological constant
\emph{Phys. Rev.} D {\bf 107} 084034, Erratum {\bf 108} 129902(E)

\bibitem{PV:2020}
Podolsk\'{y} J and Vr\'{a}tn\'{y} A
2020
Accelerating NUT black holes
\emph{Phys. Rev.} D {\bf 102} 084024

\bibitem{OPA:2025a}
Ovcharenko H, Podolsk\'{y} J and Astorino M
2025
Black holes of type D revisited: Relating their various metric forms
\emph{Phys. Rev.} D {\bf 111} 024038

\bibitem{OPA:2025b}
Ovcharenko H, Podolsk\'{y} J and Astorino M
2025
Revisiting black holes of algebraic type D with a cosmological constant
\emph{Phys. Rev.} D {\bf 111} 084016

\bibitem{FPS:2024}
Fern\'{a}ndez-\'{A}lvarez F, Podolsk\'{y} J and Senovilla J M M
2024
Analysis of gravitational radiation generated by type D black holes with positive cosmological constant
\emph{Phys. Rev.} D {\bf 110} 104029

\bibitem{PodolskyOvcharenko:2025a}
Podolsk\'{y} J and Ovcharenko H
2025
Kerr black hole in a uniform Bertotti-Robinson magnetic field: An exact solution
\emph{Phys. Rev. Lett.} {\bf 135} 181401

\bibitem{OvcharenkoPodolsky:2025b}
Ovcharenko H and Podolsk\'{y} J
2025
New class of rotating charged black holes with nonaligned electromagnetic field
\emph{Phys. Rev.} D {\bf 112} 064076

\bibitem{PlebanskiDemianski:1976}
Pleba\'{n}ski J F and Demia\'{n}ski M
1976
Rotating, charged and uniformly accelerating mass in general relativity
\emph{Ann. Phys.~(N.Y.)} {\bf 98} 98

\bibitem{Stephanietal:2003}
Stephani H, Kramer D, MacCallum M A H, Hoenselaers C and Herlt E
2003
\emph{Exact Solutions of Einstein's Field Equations}
(Cambridge: Cambridge University Press) chapter 21

\bibitem{GriffithsPodolsky:2009}
Griffiths J B and Podolsk\'{y} J
2009
\emph{Exact Space-Times in Einstein's General Relativity}
(Cambridge: Cambridge University Press) chapter 16

\bibitem{GriffithsPodolsky:2005}
Griffiths J B and Podolsk\'{y} J
2005
Accelerating and rotating black holes
\emph{Class.~Quantum Grav.} {\bf 22} 3467

\bibitem{GriffithsPodolsky:2006}
Griffiths J B and Podolsk\'{y} J
2006
A new look at the Pleba\'nski--Demia\'nski family of solutions
\emph{Int.~J. Mod.~Phys.}~D {\bf 15} 335

\bibitem{PodolskyGriffiths:2006}
Podolsk\'{y} J and Griffiths J B
2006
Accelerating Kerr--Newman black holes in (anti-)de Sitter space-time
\emph{Phys. Rev.}~D {\bf 73} 044018

\bibitem{Astorino:2024a}
Astorino M
2024
Equivalence principle and generalised accelerating black holes from binary systems
\emph{Phys. Rev.}~D {\bf 109} 084038

\bibitem{Astorino:2024b}
Astorino M
2024
Most general type-D black hole and the accelerating Reissner--Nordstrom--NUT--(A)dS solution
\emph{Phys. Rev.}~D {\bf 110} 104054

\bibitem{ChngMannStelea:2006}
Chng B, Mann R and Stelea C
2006
Accelerating Taub-NUT and Eguchi--Hanson solitons in four dimensions
\emph{Phys. Rev.}~D {\bf 74} 084031

\bibitem{Astorino:2023a}
Astorino M and Boldi G
2023
Plebanski--Demianski goes NUTs (to remove the Misner string)
\emph{JHEP} {\bf 08} 085

\bibitem{Astorino:2023b}
Astorino M
2023
Accelerating and charged type I black holes
\emph{Phys. Rev.}~D {\bf 108} 124025

\bibitem{WuWu:2024}
Wu S-Q and Wu D
2024
Is the type-D NUT C-metric really ``missing'' from the most general Plebanski--Demianski solution?
\emph{Phys. Rev}.~D {\bf 110} 104072


\bibitem{FranJose:2020}
Fern\'{a}ndez-\'{A}lvarez F and Senovilla J M M
2020
Gravitational radiation condition at infinity with a positive cosmological constant
\emph{Phys. Rev.}~D {\bf 102} 101502

\bibitem{FranJose:2022}
Fern\'{a}ndez-\'{A}lvarez F and Senovilla J M M
2022
Asymptotic structure with a positive cosmological constant
\emph{Class. Quant. Grav.} {\bf 39} 165012

\bibitem{KrtousPodolsky:2004}
Krtou\v{s} P and Podolsk\'{y} J
2004
Asymptotic directional structure of radiative fields in spacetimes with a cosmological constant
\emph{Class. Quant. Grav.} {\bf 21}  R233

\end{thebibliography}
\end{document}